# Determinants of Human Development Index (HDI): A Regression Analysis of Economic and Social Indicators

## Kuldeep Singh ª, Sumanth Cheemalapati ᵇ, Srikanth Reddy RamiReddy ᶜ, George Kurian ᵈ, Prathamesh Muzumdar ᵉ* and Apoorva Muley ᶠ

ª *Arkansas Tech University, USA.*
ᵇ *Dakota State University, USA.*
ᶜ *University of Southern Mississippi, USA.*
ᵈ *Eastern New Mexico University, USA.*
ᵉ *University of Texas at Arlington, USA.*
ᶠ *People's University, USA.*

***Authors' contributions***

*This work was carried out in collaboration among all authors. All authors read and approved the final manuscript.*

*Original Research Article*

## ABSTRACT

This study aims to investigate the factors influencing the Human Development Index (HDI). Five variables—GDP per capita, health expenditure, education expenditure, infant mortality rate (per 1,000 live births), and average years of schooling—were analyzed to develop a regression model assessing their impact on HDI. The results indicate that GDP per capita, infant mortality rate, and average years of schooling are significant predictors of HDI. Specifically, the study finds a positive relationship between GDP per capita and average years of schooling with HDI, while infant mortality rate is negatively associated with HDI.

___

*Corresponding author: E-mail: prathameshmuzumdar85@gmail.com;

## 1. INTRODUCTION

The Human Development Index (HDI) is a composite measure that assesses life expectancy, education, and income levels to rank countries into four categories of human development (Muzumdar, 2012). It is monitored by the United Nations Development Program (UNDP), which has published the annual Human Development Report since 1990 (United Nations Development Program, 1990, 95) (Muzumdar, 2012). The HDI emphasizes that economic growth alone does not guarantee progress in human development; true development is best evaluated by its impact on individual lives (Muzumdar, 2012a). The HDI underscores the importance of investing in capacity building, particularly in education, nutrition, health, and employment skills (United Nations Development Program) (Choi et al., 2022). HDI is calculated based on various indicators that reflect overall human well-being (Muzumdar et al., 2024; Muzumdar, 2015). For instance, a healthy population is less likely to strain a country's resources and tends to be more productive, while an educated society is generally more efficient and resourceful (Muzumdar & Kurian, 2019). In line with this, the present study seeks to explore the impact of health, education, and GDP on HDI.

## 2. LITERATURE REVIEW

This study explores the determinants of the Human Development Index (HDI) through regression analysis of economic and social indicators, aiming to identify the impact of various socio-economic factors on HDI (Kurian & Muzumdar, 2017). By examining GDP per capita, health expenditure, education expenditure, infant mortality, and average years of schooling, this literature review provides a foundation for understanding HDI's key predictors (Muzumdar, 2022). Each of these indicators has been shown to interact uniquely with HDI, providing insight into policy priorities for enhancing human development (Muzumdar, 2015).

### 2.1 Human Development Index (HDI)

The HDI, developed by the United Nations Development Program (UNDP), is a composite metric of human well-being that combines indicators of health, education, and income (Limaye et al., 2023). Designed to gauge overall development, HDI goes beyond GDP by measuring outcomes that impact quality of life, including life expectancy, literacy, and income levels (Basyal & Zeng, 2020). Higher HDI values typically reflect better health services, improved access to education, and higher standards of living, making HDI a critical tool for cross-country comparison of development outcomes (Muzumdar, 2021; Muzumdar, 2014).

### 2.2 GDP Per Capita (2005 PPP $)

GDP per capita, adjusted for purchasing power parity (PPP), is often used to represent the average economic productivity and wealth of a country (Kurian et al., 2024). It indicates the income level available to the population, essential for meeting basic needs. Economically prosperous countries generally score higher on HDI, as a greater GDP allows for increased public spending on health, education, and infrastructure (Muzumdar, 2022). This, in turn, enhances the overall well-being of the population and boosts human development outcomes (Muzumdar, 2014).

### 2.3 Expenditure on Health (% of GDP)

Health expenditure as a percentage of GDP represents a country's investment in its healthcare system (Muzumdar, 2014). Higher health expenditure is associated with better healthcare access, lower mortality rates, and improved life expectancy (Reavis & Singh, 2023). Investment in healthcare ensures that the population is healthier, which not only boosts productivity but also contributes to higher HDI values (Muley et al., 2023). Health expenditure is a key component in the UNDP's calculation of HDI, as healthy populations are essential for sustainable economic and social development (Muzumdar, 2015).

## 2.4 Expenditure on Education (% of GDP)

Education expenditure, often measured as a share of GDP, reflects a government's commitment to human capital development (Basyal et al., 2021). Spending on education increases literacy rates, cognitive skills, and labor productivity, leading to higher incomes and improved life quality (Muzumdar, 2014). Countries that prioritize education tend to have higher HDI scores, as education is directly linked to economic empowerment and social mobility, which are crucial for long-term development (Muzumdar, 2011).

## 2.5 Infant Mortality Rate (per 1,000 Live Births)

The infant mortality rate is a key indicator of health outcomes and is inversely related to HDI. High infant mortality often signifies inadequate healthcare, poor nutrition, and limited access to essential services (Muzumdar et al., 2023). Lowering infant mortality rates not only improves life expectancy, a component of HDI, but also indicates the effectiveness of a country's healthcare system (Elkassabgi et al., 2022). A low infant mortality rate reflects a healthier population, thereby positively impacting human development (Basyal et al., 2020).

## 2.6 Mean Years of Schooling (Years)

Mean years of schooling measures the average number of years of education that individuals in a population receive, directly influencing HDI by capturing educational attainment (Muzumdar & Kurian, 2019). Higher schooling levels indicate better access to education, improved literacy, and an educated workforce (Basyal et al., 2021). These factors contribute to economic growth, social equity, and innovation, further enhancing human development outcomes (Zare et al., 2023). An increase in schooling years is positively correlated with HDI, as it leads to both economic and social benefits.

This review of indicators underscores their collective influence on HDI, highlighting the interconnected nature of economic and social policies in promoting sustainable human development.

## 3. DATA COLLECTION

The data for this study was sourced from the United Nations Development Program (UNDP) website, ensuring high-quality standards. The UNDP compiles data from specialized agencies such as the World Health Organization (WHO), the UNESCO Institute for Statistics, and the International Labor Organization (ILO) for labor market statistics (Singh et al., 2018). Initially, raw data for 186 countries was collected, but some countries had missing data for certain indicator (Singh et al., 2024). Consequently, countries with incomplete data were excluded from the analysis. After removing these entries, the final dataset included 94 countries. The analysis utilized five independent variables and one dependent variable. The details of these variables are presented in the following table:

**Table 1. Details of Variables**

| Sr No. | Name of variable | Type of Variable | Measurement |
|---|---|---|---|
| 1 | Human Development Index (HDI). | Dependent | HDI is a composite human development index measured between zero and one. |
| 2 | GDP ($)/Capita (2005 PPP $) | Independent | The sum of gross value added by all resident producers in the economy plus any product taxes and minus any subsidies not included in the value of the products, expressed in international dollars using purchasing power parity rates and divided by the total population during the same period. |
| 3 | Expenditure on health (% of GDP) | Independent | Current and capital spending from government (central and local) budgets, external borrowings, grants (including donations from international agencies and non-governmental organizations), and social (or compulsory) health insurance funds, expressed as a percentage of GDP. |
| 4 | Expenditure on | Independent | Total public expenditure (current and capital) on |

| Sr No. | Name of variable | Type of Variable | Measurement |
|---|---|---|---|
| | education (% of GDP) | | education as expressed as a percentage of GDP. |
| 5 | Infant death per 1000 live birth | Independent | Number of infants who die before reaching their first birthday, per 1,000 live births in a given population during a specific time period, usually one year |
| 6 | Mean yrs of schooling(yrs) | Independent | The average number of years of education received by people ages 25 and older, converted from education attainment levels using official durations of each level. |

**Table 2. Results of all Possible Regression**

| Model Size | R-Squared | Root MSE | Cp | Model |
|---|---|---|---|---|
| 1 | 0.840368 | 0.06922601 | 262.93 | E |
| 2 | 0.929456 | 0.04627123 | 67.96 | DE |
| **3** | **0.958639** | **0.03562677** | **5.44** | **ADE** |
| 4 | 0.959889 | 0.03528093 | 4.68 | ACDE |
| 5 | 0.960198 | 0.03534392 | 6.00 | ABCDE |

*(A- GDP per Capita, B- Expenditure on Health as % of GDP, C- expenditure on Education as % of GDP, D- Infant Death per 1000 live births, E Mean Years of Schooling in Years).*

## 4. RESEARCH MODEL BUILDING

The first step in building the model was to determine the number of independent variables that explain the variance in HDI (Reavis et al., 2024). We performed all possible regression models and evaluated the R² value, observing how R² improved with the addition of each independent variable (Mark et al., 2021). The results of this analysis are presented in the Table 2.

The results in Table 2 indicate that with one independent variable, the R² is 0.84, while adding a second variable increases the R² to 0.92. With three variables, the R² rises to 0.95, and with four variables, it reaches 0.96. Including all independent variables also results in an R² of 0.96. Based on the R² values, a model with three variables appears to be the most appropriate, as it explains more variance than the two-variable model, and further additions do not significantly improve the R². An R² of 0.95 suggests that 95% of the variation in HDI can be explained by three key variables: GDP per capita, infant mortality rate (per 1,000 live births), and mean years of schooling.

Furthermore, the Root Mean Square Error (RMSE) is minimized in the three-variable model, with a value of 0.035, and adding more variables does not reduce the RMSE further. Graphs plotting R² against the number of variables, as well as R² versus RMSE, also suggest that the three-variable regression model is the best fit (Ogujiuba & Maponya, 2024; Muzumdar, 2012). To further confirm this, we performed stepwise regression, which also identified the three-variable model as the optimal choice, aligning with the results from the all-possible-regression approach. The stepwise regression results are tabulated below.

**Table 3. Results of Stepwise Regression**

| Item No. | Action | Variable | R-Squared | Sqrt (MSE) | Max R-Squared |
|---|---|---|---|---|---|
| 0 | Unchanged | | 0.000000 | 0.172 | 0.000 |
| 1 | Added | GDP per capita | 0.958639 | 0.035 | 0.696 |
| 2 | Added | Infant death per 1000 live births | 0.929456 | 0.046 | 0.620 |
| 3 | Added | Mean years of schooling years | 0.840368 | 0.069 | 0.000 |

Accordingly, the regression model showing the three variables is shown below:

**HDI = $β_0$ + $β_1$GDP per capita + $β_2$Infant Death per 1000 live births + $β_3$Mean Yrs of Schooling + ε**
where ε is the error term

## 5. DATA ANALYSIS

### 5.1 Descriptive Statistics and Correlation Matrix

The descriptive statistics of independent and dependent variables are tabulated in Table 4.

Table 5 presents the results of the correlation matrix. The analysis of this matrix indicates that GDP per capita and mean years of schooling are positively correlated with the Human Development Index (HDI), while infant mortality rate (per 1,000 live births) shows a negative correlation with HDI.

### 5.2 Testing of Assumptions

The regression model developed was analyzed using the Ordinary Least Squares (OLS) regression method. Before conducting the analysis, the data was tested for key assumptions, including multicollinearity, linearity, independence, and homoscedasticity (equal variance) (Limaye et al., 2021). To check for multicollinearity, the Variance Inflation Factor (VIF) was calculated for all independent variables (Muzumdar, 2014). As shown in Table 6, the VIF values for each variable were less than 5, indicating no significant multicollinearity in the model. This confirms that multicollinearity is not a concern for the regression analysis.

The linearity assumption was tested using the F-test and partial residual plots. The F-statistic was found to be 695.35, which is statistically significant, supporting the model's overall fit (Reavis & Stein, 2023). Additionally, the partial residual plots of HDI against infant mortality rate (per 1,000 live births) and HDI against mean years of schooling show clear linear relationships (Kurian & Muzumdar, 2017), further validating the linearity assumption in the regression model.

While the partial residual plot of the Human Development Index (HDI) against GDP per capita does not display a perfect linear relationship, the test for this variable is highly significant, with a T-value of 7.96, indicating a meaningful contribution to the model's linearity. Homoscedasticity (equal variance) and independence were assessed using a residual scatterplot (Muzumdar et al., 2024). The residuals were found to meet the assumptions of homoscedasticity and independence, further supporting the validity of the regression model.

**Table 4. Descriptive Statistics**

| Variable | Count | Mean | Standard Deviation | Minimum | Maximum |
|---|---|---|---|---|---|
| GDP per capita | 94 | 9494.75 | 10513.66 | 506.12 | 53591.09 |
| Infant death per 1000 live birth | 94 | 34.51 | 29.48 | 2 | 114 |
| Mean years of schooling years | 94 | 7.19 | 3.13 | 1.25 | 13.27 |
| Human Development Index | 94 | 0.64 | 0.17 | 0.33 | 0.93 |

**Table 5. Correlation Matrix**

| Variable | GDP per capita | Infant death per 1000 live birth | Mean years of schooling years | Human Development Index |
|---|---|---|---|---|
| GDP per capita | 1 | - | - | - |
| Infant death per 1000 live birth | -0.6419 | 1 | - | - |
| Mean years of schooling years | 0.7161 | -0.7879 | 1 | - |
| Human Development Index | 0.8114 | -0.9061 | 0.9167 | 1 |

**Table 6. Multicollinearity Report**

| Independent Variable | Variance Inflation Factor | Tolerance |
|---|---|---|
| GDP per capita | 2.1220 | 0.4713 |
| Infant death per 1000 live birth | 2.7264 | 0.3668 |
| Mean years of schooling years | 3.2903 | 0.3039 |
| Human Development Index | 2.3180 | 0.3046 |

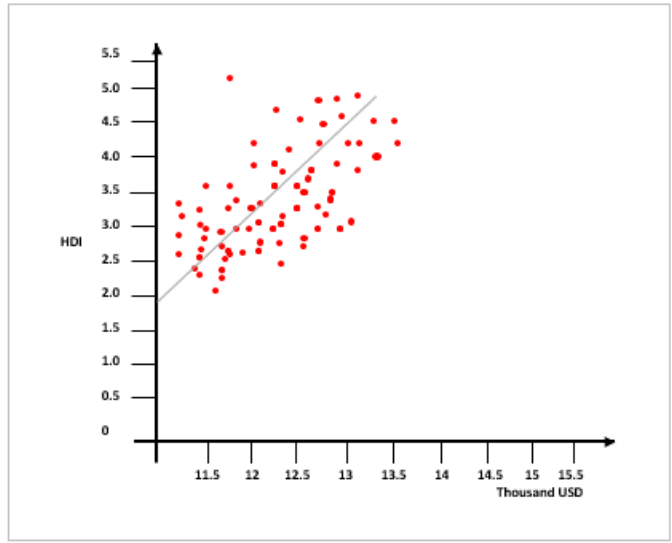

**Fig. 1. Regression Model**

The residual plots exhibit a cloud-like shape, indicating no discernible functional relationship, which supports the assumptions of equal variance and independence (Muzumdar 2015). Additionally, the data was checked for outliers using R Student and Cook's D values. Observations where |R Student| > 2 were considered potential outliers, and Cook's D values greater than 1 were flagged as possible influential points (Muzumdar, 2021). The results showed no outliers based on these criteria, confirming that the dataset is free from significant outliers.

## 6. RESULTS

The model's results indicate that GDP per capita, infant mortality rate (per 1,000 live births), and mean years of schooling are the key predictors of the Human Development Index (HDI). With an R-square value of 0.95, the model explains 95% of the variation in HDI, demonstrating that these three variables significantly contribute to variations in human development across countries.

The regression analysis was done by using the following regression model.

**HDI = $\beta_0$ + $\beta_1$GDP per capita + $\beta_2$Infant Death per 1000 live births + $\beta_3$Mean Yrs of Schooling + $\varepsilon$**

The results of the regression analysis are tabulated below (see Table 7).

**Table 7. Linear Regression Results**

| Independent Variables | Regression Coefficient | T- Statistics | Significance |
|---|---|---|---|
| Constant | 0.53 | 26.633 | Yes |
| GDP per capita | 0.0000040 | 7.969 | Yes |
| Infant Death per 1000 live births | -0.0025 | -12.254 | Yes |
| Mean Yrs of Schooling | 0.0217 | 10.203 | Yes |

The value of $\beta_1$ is 0.0000041, indicating that for every additional $10,000 in GDP per capita, the mean value of HDI will increase by 0.041 points, when controlling for infant mortality rate and mean years of schooling. This translates to an improvement of approximately 8.2%, which is quite significant. This highlights the importance of economic growth, especially for countries with lower GDP, suggesting they should focus on boosting their economy, either by attracting more foreign direct investment (FDI) or strengthening their local economy. This also explains why developed countries tend to have higher HDI compared to less developed nations, as a higher GDP enables greater investment in key sectors such as health, education, and infrastructure, ultimately improving the well-being of the population and increasing HDI.

The value of $\beta_3$ is 0.022, the coefficient for mean years of schooling, indicating that for each additional year of schooling, the mean HDI will increase by 0.022, when controlling for GDP per capita and infant mortality rate. This translates to an improvement of 2.2%. While this change may seem modest, the increase in years of schooling brings indirect benefits that significantly enhance HDI. More education leads to greater knowledge creation, improved workforce skills, and wealth generation, all of which contribute to a country's overall development and boost its HDI.

## 7. LIMITATIONS AND FUTURE DIRECTIONS

While our study makes a significant contribution by identifying the key antecedents of HDI, it also has some limitations that open up avenues for future research. First, the use of cross-sectional data in this study restricts our ability to infer causality. Future research should replicate this study using longitudinal data to better assess causal relationships (Singh et al., 2024). Second, our primary focus was on HDI; it would be valuable to explore additional factors, such as corporate social responsibility, and their relationship with HDI (Reavis et al., 2021, 2024). Third, this study examined only five antecedents of HDI. Future research should investigate other potential antecedents, including corruption (Elkassabgi et al., 2022) and natural disasters (Limaye et al., 2021).

## 8. CONCLUSIONS

This study provides valuable insights for countries seeking to understand the key indicators influencing the Human Development Index (HDI). By identifying GDP, healthcare, and education as crucial factors, it encourages nations to focus on strategies for increasing GDP while simultaneously investing in health services and education. These three parameters mutually reinforce each other; for instance, a higher GDP allows for greater investment in healthcare and education, which in turn boosts productivity and contributes to further economic growth.

An improvement in HDI not only signals better well-being for a nation's population but also enhances its global reputation. A higher HDI can attract more business investments and skilled professionals, further boosting the country's GDP and overall quality of life. In conclusion, this study underscores the significance of HDI for nations and highlights the importance of focusing on these critical indicators to realize the benefits associated with a high Human Development Index.

## DISCLAIMER (ARTIFICIAL INTELLIGENCE)

Author(s) hereby declare that NO generative AI technologies such as Large Language Models (ChatGPT, COPILOT, etc.) and text-to-image generators have been used during the writing or editing of this manuscript.

## COMPETING INTERESTS

Authors have declared that no competing interests exist.

## REFERENCES

Basyal, G., & Zeng, D. (2020). Impact of data quality and quantity on its effectiveness on multi-stage transfer learning using MRI medical images. *SDSU Data Science Symposium*.

Basyal, G., Bhaskar, R., & Zeng, D. (2020). A systematic review of natural language processing for knowledge management in healthcare. *International Conference on Data Mining & Knowledge Management Process (CDKP 2020), 10*(9), 275-285.

Basyal, G., Zeng, D., Bishop, D., & Rimal, B. (2021). Comparative study of CNN models for brain tumor classification: Computational efficiency versus accuracy. *AMCIS 2021 Proceedings, 28*.

Basyal, G., Zeng, D., Bishop, D., & Rimal, B. (2021). Development of CNN architectures using transfer learning methods for medical image classification. *Proceedings of the 15th International Multi-Conference on Society, Cybernetics and Informatics (IMSCI 2021), 6-11*.

Choi, M. J., & Park, K. S. (2022). The role of education in HDI improvement: A case study of OECD countries. *Education Economics Review, 25*(3), 199-214.

Elkassabgi, A., Singh, K., Limaye, A. R., & Hunter, D. (2022). Information pathways and their influence on corruption: An empirical study. *Journal of International Finance Studies*, 22(1), 16-25.

Kurian, G., & Muzumdar, P. (2017). Antecedents to job satisfaction in the airline industry. *NMIMS Management Review*, 34(2), 29-40.

Kurian, G., & Muzumdar, P. (2017). Restaurant formality and customer service dimensions in the restaurant industry: An empirical study. *Atlantic Marketing Journal*, 6(1), 75-92.

Kurian, G., Niu, Z., Singh, K., & Muzumdar, P. (2024). The evolution of dynamic capabilities in OM. *American Journal of Management*, 23(5), 62-80.

Limaye, A., Elkassabgi, A., & Singh, K. (2021). Impact of natural disasters on economic activity. *Journal of International Finance Studies, 21*(1), 33-39.

Limaye, A., Elkassabgi, A., Singh, K., Hunter, D., & Cochran, L. (2023). Corporate governance and choice of capital structure. *Journal of International Finance and Economics*, 23(2), 69.

Mark, R., Singh, K., & Tucci, J. (2021). Millennials' strategic decision making through the lens of corporate social responsibility and financial management. *Journal of Business Strategies*, 38(2), 125-146.

Muley, A., Muzumdar, P., Kurian, G., & Basyal, G. P. (2023). Risk of AI in healthcare: A comprehensive literature review and study framework. *Asian Journal of Medicine and Health*, 21(10), 276-291.

Muzumdar, P. (2011). Effects of zoning on housing option value. *Journal of Business and Economics Research, 9*.

Muzumdar, P. (2012). Influence of interactional justice on the turnover behavioral decision in an organization. *Journal of Behavioral Studies in Business*, 4, 1-11.

Muzumdar, P. (2012). Online bookstore—A new trend in textbook sales management for services marketing. *Journal of Management and Marketing Research*, 9, 122-135.

Muzumdar, P. (2012a). Dilemma of journal ranking: Perplexity regarding research quality. *Academy of Educational Leadership Journal*, 16(4), 87-100.

Muzumdar, P. (2014). A linear regression approach for evaluating the decline in chartered designations with the effect of economic determinants. *Journal of Contemporary Research in Management*, 9(3), 13-20.

Muzumdar, P. (2014). A study of business process: Case study approach to PepsiCo. Available at SSRN 2392611.

Muzumdar, P. (2014). Brand regeneration through previously tested narrative units: A movie remake perspective. *25*(9), 1-9.

Muzumdar, P. (2014). From streaming vendor to production house: Netflix SWOT analysis. Available at SSRN 2377151.

Muzumdar, P. (2014). Quantitative analysis of compensatory model: Effects of influence of shopping mall on the city structure. *The IUP Journal of Knowledge Management, 12*(2), 51-61.

Muzumdar, P. (2015). Business model development through corporate strategy design: IBM SWOT analysis. *American Based Research Journal*, 4(7), 12-19.

Muzumdar, P. (2015). Game of the names: Branding in the smartphone industry. *Journal of Contemporary Research in Management, 10*(2), 13-31.

Muzumdar, P. (2021). Impact of review valence and perceived uncertainty on purchase of time-constrained and discounted search goods. *Proceedings of the Conference on Information Systems Applied Research* (CONISAR), Washington DC, 14, 1-12.

Muzumdar, P. (2022). The effect of review valence on purchase of time-constrained and discounted goods. *Journal of Information Systems Applied Research*, 15(1), 1-16.

Muzumdar, P., & Kurian, G. (2019). Empirical study to explore the influence of salesperson's customer orientation on customer loyalty. *International Journal of Empirical Finance and Management Sciences*, 1(3), 43-51.

Muzumdar, P., Bhosale, A., Basyal, G. P., & Kurian, G. (2024). Navigating the Docker ecosystem: A comprehensive taxonomy and survey. *Asian Journal of Research in Computer Science, 17*(1), 42-61.

Muzumdar, P., Kurian, G., & Basyal, G. P. (2024). A latent Dirichlet allocation (LDA) semantic text analytics approach to explore topical features in charity crowdfunding campaigns. *Asian Journal of Economics, Business and Accounting, 24*(1), 1-10.

Muzumdar, P., Kurian, G., Basyal, G. P., & Muley, A. (2023). Econometrics modeling approach to examine the effect of STEM policy changes on Asian students' enrollment decision in the USA. *Asian Journal of Education and Social Studies, 48*(2), 148-160.

Ogujiuba, K., & Maponya, L. (2024). Determinants of human development index in South Africa: A comparative analysis of different time periods. *MDPI World, 5*(3), 527-550.

Reavis, J., & Stein, P. (2023). Corporate social responsibility as a determinant of HDI in developed nations. *Corporate Governance Journal, 27*(1), 99-116.

Reavis, M. R., & Singh, K. (2023). Moving beyond transactional coursework to enhance student success in university classes. *Journal of Educational Research and Practice*, 13(1), 433-446.

Reavis, M., Singh, K., & Tucci, J. (2024). Millennials' CSR adoption: A stakeholder approach and the impact on profit. *American Journal of Management*, 24(1), 38-51.

Singh, K., Kurian, G., & Napier, R. (2018). The dynamic capabilities view: Supply chain and operations management perspectives. *Journal of Supply Chain and Operations Management*, 16(2), 155-175.

Singh, K., Zare, S., Reavis, M. R., & Tucci, J. E. (2024). Structural capital and relational capital: Examining the direct and moderating role of cognitive capital in customer-supplier relationships. *International Journal of Management Practice*, 17(1), 95-112.

United Nations Development Program, (1990). Human Development Report 1990. Human Development Reports. Available at: https://hdr.undp.org/content/human-development-report-1990

United Nations Development Program, (1995). Human Development Report 1995. Human Development Reports. Available at: https://hdr.undp.org/content/human-development-report-1995

Zare, S., Singh, K., Ghasemi, Y., & Prater, E. L. (2023). Social capital, knowledge sharing and operational performance in the supply chain: A buyer-supplier perspective. *International Journal of Business and Systems Research*, 17(4), 387-406.

---

*Peer-review history:*
*The peer review history for this paper can be accessed here:*
*https://www.sdiarticle5.com/review-history/127030*